\documentclass[onecolumn,prd,showpacs,preprintnumbers,amsmath,amssymb,floatfix]{revtex4-2}

\usepackage{graphicx}

\usepackage{bm}
\usepackage{amsfonts}
\usepackage{lineno,hyperref}
\usepackage{array}
\usepackage{microtype}
\usepackage{float}

\begin{document}
	
\title{Gravastar under the framework of braneworld gravity II: effect of the Kuchowicz metric function}

\author{Saibal Ray}
\email{saibal@associates.iucaa.in}
\affiliation{Department of Physics, Government College of Engineering and Ceramic Technology, Kolkata 700010, West Bengal, India}

\author{Shounak Ghosh\footnote{Corresponding author}}
\email{shounakphysics@gmail.com}
\address{Directorate of Legal Metrology, Department of Consumer Affairs, Govt. of West Bengal, Alipurduar 736121, West Bengal, India }
	
\author{Rikpratik Sengupta}
\email{rikpratik.sengupta@gmail.com}
\address{Department of Physics, Government College of Engineering and Ceramic Technology, Kolkata 700010, West Bengal, India.}

\date{\today}

\begin{abstract}
In recent years, a class of compact objects called gravastars have drawn immense interest as regular solutions to end state stellar collapse. Since the energy density involved in collapse process is expected to be high, it is a natural choice to study such compact objects in context of modified gravity theories which reduce to General Relativity (GR) in the low energy regime. We have already framed a model of gravastar in such a modified gravity framework involving extra dimensional Randall-Sundrum (RS) single brane gravity~[Phys. Rev. D \textbf{102}, 024037 (2020)]~\cite{Sengupta1}. As a sequel in the present paper we substantially improve our previous model by choosing the Kuchowicz function as one of the metric potentials, which leads to many new interesting results and physical features from our analysis as discussed in this paper. Also, we provide essential additional stability checks on our gravastar model to investigate the possibility of any instability creeping in due to the higher dimensional framework. Our present improved gravastar model is found to clear all the stability checks successfully. Very interestingly, the static spherically symmetric matter distributions are found to accommodate both classes of solutions obeying and violating the modified energy conditions on the RS brane as we find in this work. We can conclude from our analysis that the Kuchowicz metric potential is very effective for describing regular solutions to compact objects at substantially high energies on the 3-brane.	
\end{abstract}


\maketitle

\section{Introduction} \label{sec1}

Compact stellar objects like white dwarfs, neutron stars (more recently strange stars) and black holes have been studied extensively in literature~\cite{Schatzman,Liebert,Henry,Glendenning,Baym,Xu,Schaffner-Bielich,Misner,Wald,Carroll,Townsend,Luminet}. Black hole as the final state of stellar collapse has been a matter of serious research interest for general relativists since many decades~\cite{Hawking1}. Also, due to the geometric approach that must be taken to comprehensively analyse a black hole, the presence of singularity in the solutions and the extremely high levels of energy densities involved in black holes due to its compactness, black hole physics provides a very good testbed for theories of quantum gravity~\cite{Agullo,Ashtekar2,Wald1,Susskind1} and gravity theories which reduce to general relativity (GR) at the low energy limit~\cite{Casodio,Dadhich1,Deruelle,Germani,Visser}. In general, there is scope for modification in the Einstein Field Equations (EFE's) in both the geometrical and matter sectors. In modified gravity theories, the geometrical modification is in many cases expressed as an effective modification in the matter sector. As the contribution due to the additional physical parameters are made to vanish in the modified EFEs,we get back the GR field equations. However, direct modifications in the matter source does not modify the gravity theory but adds new matter-energy ingredients within the standard GR framework. In quantum gravity theories, it is attempted to quantize the gravitaional sector~\cite{Ashtekar,Rovelli} in order to bring it in equal footing with the matter sector which can be well quantized within the standard model of particle physics.

As in the case of the initial singularity of the universe, some cosmologists are unhappy about the presence of singularity and consider this a consequence of the incompleteness of GR~\cite{Borde1,Borde2}. Many alternative scenarios to the initial big bang singularity have been put forward modifying both the geometry and matter sectors of the standard EFEs, among which two major ones are the emergent\cite{Ellis1,Ellis2,bcp,banerjee,sahni,mulryne,cai} and bouncing~\cite{shtanov,singh,bojowald,corichi} universes. Similarly in black hole physics, any singularity occuring at the event horizon can be removed by a co-ordinate transformation \cite{Finkelstein,Eddington,Kruskal,Szekeres} but the singularity at the centre is a curvature singularity where the Riemann tensor diverges and cannot be removed by co-ordinate transfromation~\cite{Hawking1,Hawking2}. So, there have been a number of proposed regular solutions in this context also~\cite{Banados1,Banados2,Markov,Polchinski,Bardeen}. The gravastar or gravitaionally vacuum condensate star is one such regular solution involving end state stellar collapse, which had been proposed by Mazur and Mottola at the beginning of the century~\cite{Mazur2001,Mazur2004}. The central idea is to replace the interior of the collapsing star by a form of gravitaional Bose-Einstein condensate, which taking quantum effects into consideration, replaces the horizon by a critical surface of a gravitational phase transition~\cite{Chapline1,Laughlin,Chapline2}. Such an interior can be well described by the equation of state (EOS) $p=-\rho$~\cite{Gliner}, thus leading to a repulsive, cosmological constant like effect which prevents gravtional collapse of the stellar object to infintesimal microscopic volume, in order,preventing the energy density and curvature from diverging and not favouring formation of central singularity. Inside the shell, there is stiff matter whose presence was first proposed by Zeldovich~\cite{Zeldo1,Zeldo2}, and is described by the EOS $p=\rho$.

It was first proposed by Randall and Sundrum (RS) in 1999~\cite{Randall1,Randall2} that we live on a (3+1)-hypersurface embedded in a higher dimensional space-time called the bulk. The forces and particles involved in the standard model of particle physics are confined to this hypersurface while gravity is free to propagate in the bulk, letting us feel its extra dimensional manifestations on the 3-brane. The single brane RS scenario has been extremely popular in addressing cosmological and astrophysical problems and have led to plenty of interesting results over the last two decades~\cite{Binetruy,Maeda,Maartens,Langlois,Chen,Kiritsis,Shiromizu,Campos,Germani,Deruelle,Wiseman1,Visser,Creek,Pal,Dadhich,Bruni,Govender,Wiseman2,Ovalle5,Ovalle6}. Previously we have studied static, spherically symmetric gravastar~\cite{Sengupta1} and wormhole solutions~\cite{Sengupta2} on the brane. Wormhole solution with the Kuchowicz function as one of the metric potentials have led to many interesting consequences on the brane, mainly imposing constraints on the brane tension which is one of the most important parameters in the context of RS gravity. As the brane tension tends to infinitely large values, the energy scales correspondingly reduce to smaller values thus making the terms encoding the corrections to standard GR redundant. Fundamental and interesting constraints were found to be imposed on brane gravity (BG) from the wormhole geometry. As gravastars are believed to have high energy densities, it would be interesting to analyze their behaviour in the BG framework imposing the Kuchowicz metric function as one of the metric potential. Gravastars have also been explored in other modified gravity theories~\cite{Shamir,Das,Debnath2019a,Debnath2019b}. A comprehensive review on gravastar may be found in~\cite{Ray2020}.

First we lay down a brief review of the mathematical framework of BG. Then we consider the solutions of the modified EFEs in the interior and shell regions employing the thin-shell approximation for the gravastar. In the following section we make an in-depth analysis of the different physical properties of the gravastar including the matter density, entropy, energy and the shell thickness followed by matching of the solutions at the interior-shell interface. Invoking the Israel-Darmois junction conditions, we obtain the surface density and surface pressure for the shell. The variations of all the physical properties have been done as function of the radial distance from the interior to the exterior shell surface. Finally we impose a three-fold stability check on our gravastar model on the brane from the surface redshift, energy condition and cracking condition viewpoint. We conclude with a detailed discussion of the interesting consequences and inferences that can be drawn from our analysis.

\section{Mathematical formalism and solutions} \label{sec2}

First we briefly review the mathematical framework that is the modified EFEs on the 3-brane. In the brane we do not introduce any form of matter that can modify the stress energy tensor, but due to the physical or rather purely geometrical implications we get some correction terms which gives an effective matter description on the 3-brane. The EFE reads
\begin{equation} \label{eq1}
G_{\mu\nu}=T_{\mu\nu}+\frac{6}{\sigma} S_{\mu\nu}-E_{\mu\nu},
\end{equation}
where $\sigma$ represents the brane tension. Here we use natural units. The BG idea introduces corrections at higher energies, which comprises of quadratic terms in the energy momentum tensor ($S_{{\mu}{\nu}}$) and also projection of the Weyl tensor on the 3-brane ($E_{{\mu}{\nu}}$). The correction terms can be expressed as
\begin{equation} \label{eq2}
S_{\mu\nu}= \frac{TT_{\mu\nu}}{12}-\frac{T_{\mu\alpha}T_{\nu}^{\alpha}}{4}+ \frac{g_{\mu\nu}}{24}(3T_{\alpha\beta}T^{\alpha\beta}-T^2),
\end{equation}

\begin{equation} \label{eq3}
E_{\mu\nu}=-\frac{6}{\sigma}[Uu_{\mu} u_{\nu}+Pr_{\mu}r_{\nu}+h_{\mu\nu}(\frac{U-P}{3})],
\end{equation}
where $T$ is the trace of the stress-energy tensor, $u_{\mu}$ gives 4-velocity and $r_{\mu}$ is radial vector on the brane. $U$ and $P$ respectively represents the energy density and pressure of the 5-dimensional embedding space. They can be written by the EOS $P=\omega U$~\cite{Castro},such that  $-3 < \omega < 2$ is the range of the EOS parameter.

The stress-energy tensor on the brane reads as
\begin{equation} \label{eq4}
T_{\mu\nu}=\rho u_{\mu} u_{\nu}+ ph_{\mu\nu},
\end{equation}
where $h_{\mu\nu}=g_{\mu\nu}+u_{\mu} u_{\nu}$ is the projection of the 5-dimensional metric on the brane and $U=A\rho+C$~\cite{Banerjee2}, $A$ and $C$ being constant model parameters to be estimated later.

There is no modification in the conservation equation which is written as
\begin{equation} \label{eq5}
	\frac{dp}{dr}=-\frac{1}{2}\frac{d\nu}{dr}(p+\rho).
\end{equation}

For studying gravastar solutions, we need to obtain the solutions for the modified EFEs on the brane in the case of a static, spherically symmetric line-element written as
\begin{equation}\label{eq6}
ds^2=-e^{\nu(r)}dt^2+e^{\lambda(r)}dr^2+r^2(d\theta^2+sin^2\theta d\phi^2).
\end{equation}

From Eq. \ref{eq1} we get
\begin{eqnarray}
&&e^{-\lambda}\left(\frac{\lambda'}{r}-\frac{1}{r^2}\right)+\frac{1}{r^2} =\rho^{eff},\label{eq7}\\
&&e^{-\lambda}\left(\frac{\nu'}{r}+\frac{1}{r^2}\right) -\frac{1}{r^2} =p_r^{eff},\label{eq8}\\
&&e^{-\lambda}\left(\frac{\nu''}{2}-\frac{\lambda' \nu'}{4}+\frac{\nu'^2}{4}+\frac{\nu'-\lambda'}{2r}\right) = p_t^{eff}.\label{eq9}
\end{eqnarray}

The effective energy density and effective radial and tangential pressures on the brane can be expressed as
\begin{eqnarray}\nonumber
&&\rho^{eff}=\rho(r) \left( 1+\frac {\rho(r) }{2 \sigma} \right) +{\frac {6 U}{\sigma}},\\ \nonumber
&&p_r^{eff}=p(r)+{\frac {\rho (r)  \left( p(r)+\frac{\rho(r)}{2} \right)} {\sigma}}+{\frac {2U}{\sigma}}+{\frac {4 P}{\sigma}},\\ \nonumber
&&p_t^{eff}=p(r)+{\frac{\rho(r) \left(p(r)+\frac{\rho(r)}{2}\right)}{\sigma}}+{\frac{2U}{\sigma}}-{\frac {{2 P}}{\sigma}}.\nonumber
\end{eqnarray}

From the above expressions we can clearly see that there is an in-built pressure anisotropy $6P/{\sigma}$ present in the EFE, which justifies Cattoen's\cite{Cattoen} claim of pressure anisotropy being an inherent feature of gravastars which we do not find in the relativistic framework but appears intrinsically on the brane.

The metric potential $e^{\nu(r)}$  is given by the non-singular Kuchowicz function~\cite{Kuchowicz}
\begin{equation}
	e^{\nu(r)}=e^{Br^2+2\ln K},
\end{equation}
$B$ and $K$ being constants such that $B$ has inevrse length squared dimension and $K$ is dimensionless.

\subsection{Interior solution}

We know the EOS for the interior and on using it in Equation (5), the pressure and energy density of the interior turns out to be constant such that the energy density is denoted by $\rho_c$. If we make use of this in the EFE, employing the Kuchowicz function as the first metric potential, the unknown metric potential for the interior takes the form
\begin{equation} \label{eq10}
{{\rm e}^{-\lambda (r) }}=  -\frac{8 \pi   \rho_c}{3} \left( {\frac {2 \sigma+ \rho_c}{2\sigma}} \right) {r}^{2}-\frac{16 \pi r^2}{k^4 \sigma}  \left( A \rho_c+C \right) +1+\frac{C_1}{r},
\end{equation}
where the constant of integration is put $C_1=0$ in order to preserve the regularity of the metric potential at $r=0$. On simplifying, we have
\begin{equation} \label{eq11}
{{\rm e}^{-\lambda (r) }}=-\frac{8r^2\pi   \rho_c}{3}  \left( {\frac {2 \sigma+ \rho_c}{2\sigma}} \right) -{\frac {16 \pi \left( A \rho_c+C \right) r^2}{k^4\sigma}}+1
\end{equation}

\subsection{Active gravitational mass $M(R)$}
The active gravitational mass has the form
	\begin{equation} \label{eq15}
	M(R)= 4\pi\int_{0}^{R}\rho^{eff}r^2dr=\frac{32}{3} \pi^2 R^3   \left(  \rho_c \left( 1+{\frac { \rho_c}{2\sigma}} \right) + {\frac {6(A \rho_c+C)}{{k}^{4}\sigma}} \right),
	\end{equation}
where a dependence on the brane tension may be noted.

\subsection{Intermediate thin Shell}

   From the seminal work~\cite{Mazur2001,Mazur2004}, we come to know that if we donot consider the shell to be thin, then it will not be possible to find analytical solutions of the field equations in this region, which is true even in the context of brane gravity as we see in~\cite{Sengupta1}. ALso, it is true that the thickness is directly proportional to the metric function $e^{-\lambda}$ as a result of which approximation for thin shell of finite thickness means $e^{-\lambda}\ll 1$~\cite{Mazur2001,Mazur2004}. So, the field equations take the form
\begin{equation} \label{eq16}
\frac{1}{r^2}+ \left( {\frac {\lambda' }{r}}-\frac{1}{r^2} \right) {{\rm e}^{-\lambda (r) }}
=8\pi  \left( \rho \left( 1+{\frac {\rho}{2\sigma}} \right) +{\frac {6(A\rho+C)}{{k}^{4}\sigma}} \right),
\end{equation}

\begin{equation} \label{eq17}
-\frac{1}{r^2}+ \left( 2 B+\frac{1}{r^2} \right) {{\rm e}^{-\lambda (r) }}=8\pi  \left( p+{\frac {\rho \left( p+\frac{\rho}{2} \right) }{\sigma}}+{\frac {2(A\rho+C)}{{k}^{4}\sigma}}+{\frac {4\omega \left( A\rho+C \right) }{{k}^{4}\sigma}} \right),
\end{equation}

\begin{equation} \label{eq18}
\frac{{{\rm e}^{-\lambda (r) }}}{4} \left( 8 B+4 {B}^{2}r^2-2 Br\lambda' -{\frac{2\lambda' }{r}} \right) =8\pi  \left( p+{\frac {\rho \left( p+\frac{\rho}{2} \right) }{\sigma}}+{\frac {2(A\rho+C)}{{k}^{4}\sigma}}-{\frac {2\omega \left( A\rho+C \right) }{{k}^{4}\sigma}} \right).
\end{equation}

We can find the solution for the metric potential to be given as
\begin{equation} \label{eq19}
{{\rm e}^{-\lambda (r) }}={\frac {12  e^{-2Br^2}\pi  k^4r^2{\rho_0}^{2}+8 \rho_0\pi  r^2 \left( k^4\sigma+2A \left( \omega+\frac{1}{2} \right) \right) {{\rm e}^{-Br^2}}+32 C\pi   \left( \omega+\frac{1}{2} \right) r^2+k^4\sigma}{k^4\sigma  \left( 2 Br^2+1 \right) }}.
\end{equation}

The solution depends on both the brane tension and the bulk EOS parameter. In the case of the interior solution, the metric potential $\lambda$ was found to depend on the brane tension only. So, the additional dependence on the bulk EOS parameter is characteristic of the thin shell.

\section{Physical property}

It is essential to analyze the physical properties of the shell of the gravastar. The key physical properties include the effective energy density and pressures in the radial and tangential directions of the matter composing the shell, the energy entropy and proper thickness of the shell.

\subsection{Matter density and pressure of the shell}

Using the EOS for shell matter in Eq. (5), the energy density inside the shell may be expressed as
\begin{equation}\label{eq19}
\rho=\rho_0 e^{-\nu}.
\end{equation}

So the effective matter density can be calculated to be
\begin{equation}\label{eq19}
\rho_{eff}=8 \pi \left( \rho_0 {{\rm e}^{-Br^2}}+{\frac {{\rho_0}^{2} e^{-2Br^2}}{2\sigma}}+{\frac {6(A\rho_0 {{\rm e}^{-Br^2}}+C)}{k^4\sigma}} \right).
\end{equation}

The effective radial pressure  can be calculated as
\begin{equation}\label{eq19}
p_{reff}=8 \pi \left( \rho_0 {{\rm e}^{-Br^2}}+{\frac {3{\rho_0}^{2} e^{-2Br^2}}{2\sigma}}+ {\frac {2(A\rho_0 {{\rm e}^{-Br^2}}+C)}{k^4\sigma}}+ {\frac {4\omega  \left(A\rho_0 {{\rm e}^{-Br^2}}+C \right) }{k^4\sigma}} \right).
\end{equation}

The effective tangential pressure  can be calculated as
\begin{equation}\label{eq19}
p_{teff}=8 \pi   \left( \rho_0 {{\rm e}^{-Br^2}}+{\frac {3{\rho_0}^{2}  {{\rm e}^{-2Br^2}}  }{2\sigma}}+ {\frac {2(A\rho_0 {{\rm e}^{-Br^2}}+C)}{k^4\sigma}}- {\frac {2\omega  \left(A\rho_0 {{\rm e}^{-Br^2}}+C \right) }{k^4\sigma}} \right).
\end{equation}

\begin{figure}[!htp]
\centering
\includegraphics[width=5cm]{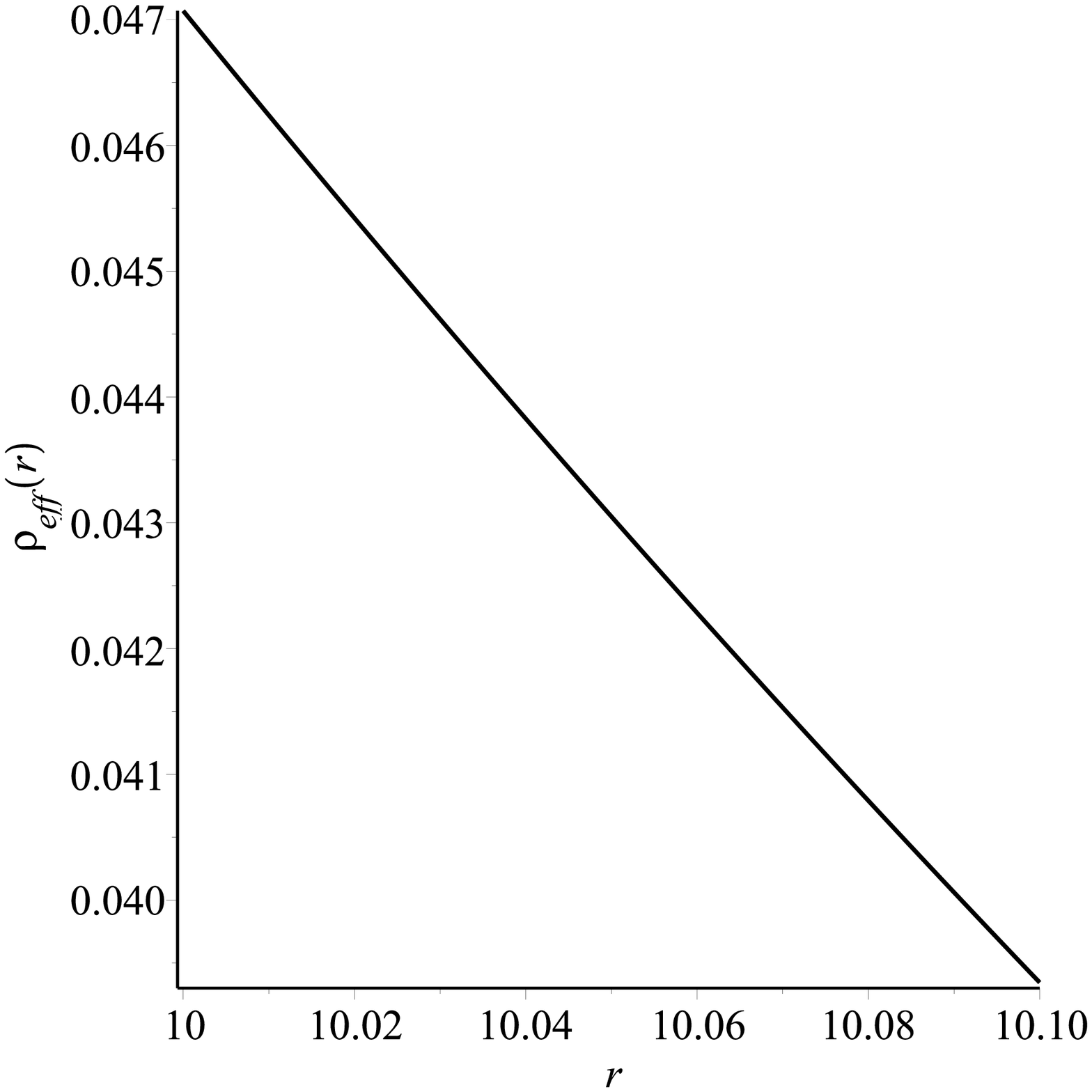}
\includegraphics[width=5cm]{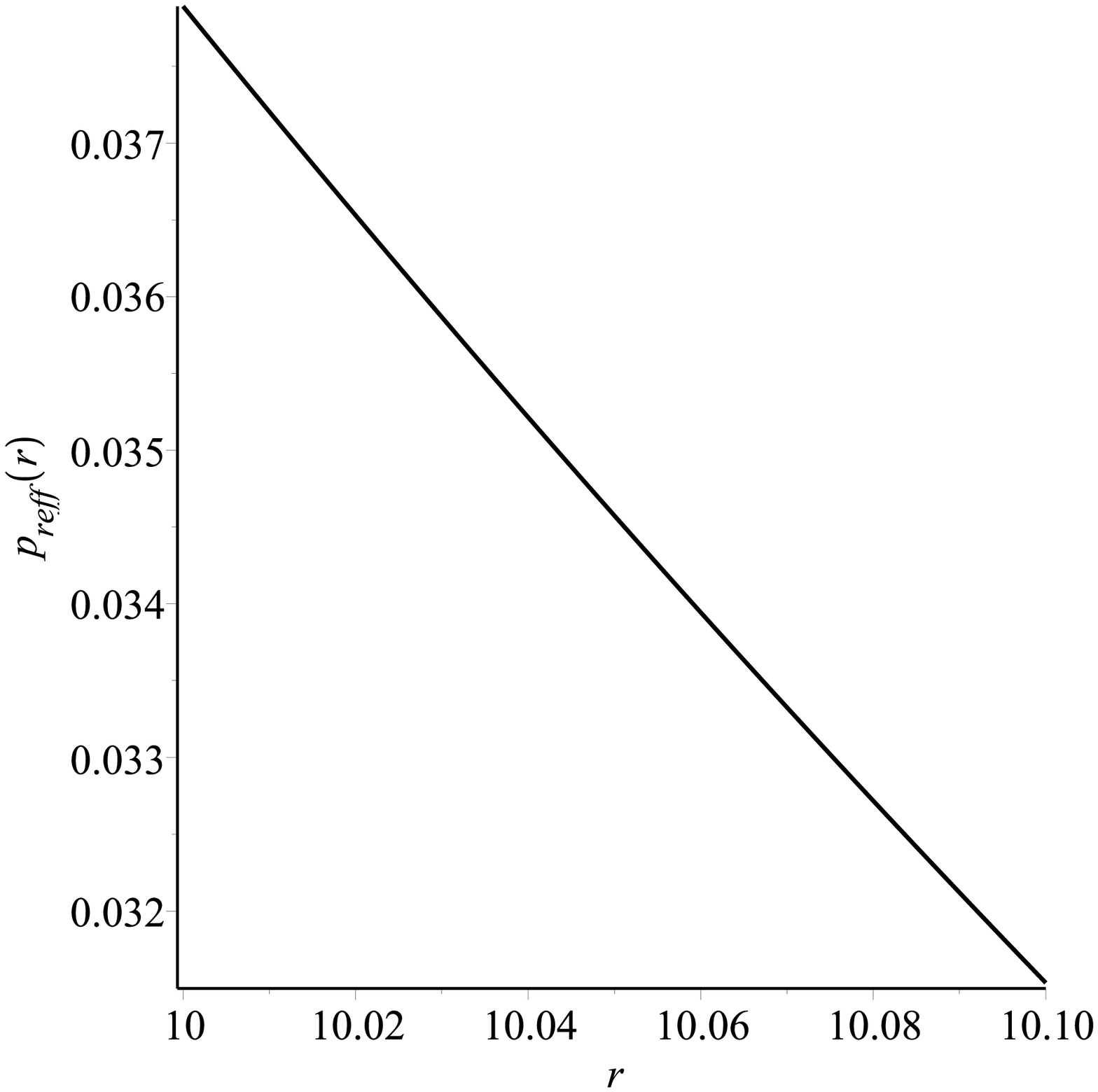}
\includegraphics[width=5cm]{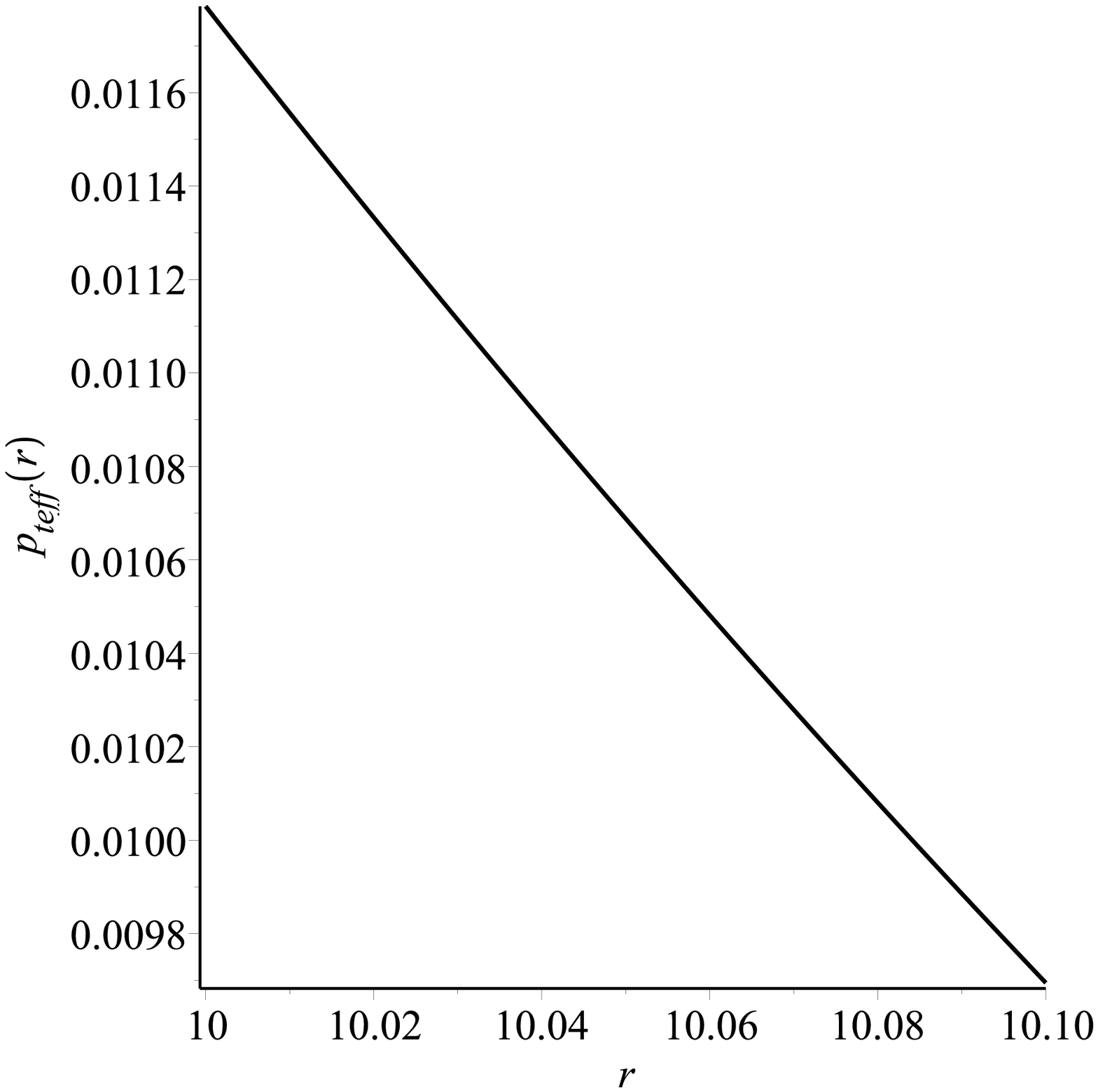}
\caption{Variation of the effective density, effective radial pressure and effective tangential pressure w.r.t. the radial coordinate $r$ for different
strange star candidates.}\label{pres.}
\end{figure}

As we can see from our obtained results the effective energy density depends on the brane tension and the effective pressures depends on both brane tension and the EOS parameter in the bulk. In the low energy limit, the brane tension $\sigma \rightarrow \infty$, making the contribution of the terms containing higher dimensional correction redundant, giving the results we would have obtained in the framework of GR. The variation of the effective density, radial and tangential pressures along the radial distance from the shell-interior interface have been shown in Fig. 1. We see that all the three parameters is maximum at the shell-interior interface and monotonically decrease in a linear pattern as we move towards the shell-exterior interface.

\subsection{Energy}

The energy of the thin shell can be obtained by integrating the effective matter density along the radial expanse of the shell, on doing which we obtain
\begin{eqnarray}\label{eq19}
E &=& 4\pi \int_{R}^{R+\epsilon}\rho^{eff}r^2 dr  \nonumber\\
E&=& \left[16 {\pi }^{2} \left( 2 \rho_0  \left( {\frac {r{{\rm e}^{-Br^2}}}{2B}}+{\frac {\sqrt {\pi }{\rm erf} \left(\sqrt {B}r\right)}{4{B}^{\frac{3}{2}}}} \right) +{\frac {{\rho_0}^{2}}{\sigma} \left( {\frac {r{{\rm e}^{-2 Br^2}}}{4B}}+{\frac {\sqrt {\pi }\sqrt{2}{\rm erf} \left(\sqrt {2}\sqrt {B}r\right)}{16{B}^{\frac{3}{2}}}} \right) }\right.\right.\nonumber\\
&+&\left.\left. {\frac {12A\rho_0}{k^4\sigma} \left({\frac {r{{\rm e}^{-Br^2}}}{2B}}+{\frac {\sqrt {\pi }{\rm erf} \left(\sqrt {B}r\right)}{4{B}^{\frac{3}{2}}}} \right) }+{\frac {4{r}^{3}C}{k^4\sigma}} \right)\right]_R^{R+\epsilon}.
\end{eqnarray}

\begin{figure*}[thbp]
\centering
\includegraphics[width=5cm]{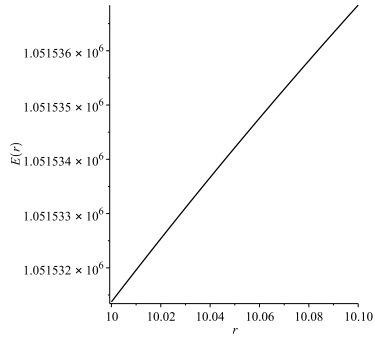}
\caption{Variation of the energy of the shell with respect to $r$.}
\end{figure*}

As we can see in Fig. 2, the energy distribution is continuous across the shell and monotonaically increases in a linear pattern as we move from the interior to the exterior. Also, from Eq. (22) we find that the energy depends on the brane tension.

\subsection{Entropy}
In order to get a measure of the entropy of the shell we use the equation
\begin{equation}\label{eq28}
S=\int_R^{R+\epsilon} 4 \pi r^2 s(r) \sqrt{e^{\lambda}} dr,
\end{equation}
where the entropy density $s(r)$ is given by
\begin{equation}\label{eq29}
 s(r)=\frac{\xi^2 k_B^2 T(r)}{4\pi\hbar^2}=\frac{\xi k_B}{\hbar}\sqrt{\frac{p}{2\pi}},
\end{equation}
 $\xi$ being constant having no dimension. In natural units, the entropy density reduces to
 \begin{equation}\label{eq30}
 s(r)=\xi\sqrt{\frac{p}{2\pi}}.
\end{equation}

The shell entropy for the brane gravastar in our case turns out to be
\begin{eqnarray}\label{eq31}
S&=& 4\pi \xi \int_{R}^{R+\epsilon} r^2 \sqrt{\frac{p e^\lambda}{2\pi}}dr \nonumber\\
=&&{\frac { 2\sqrt{2\pi }\alpha k_{{B}}\epsilon r^2}{h} \sqrt{{\frac {\rho_0{{\rm e}^{-Br^2}}k^4\sigma  \left( 2 Br^2+1 \right) }{32 \rho_0\pi  r^2 \left( \frac{k^4\sigma}{4}+A \left( \omega+\frac{1}{2} \right)  \right) {{\rm e}^{-Br^2}}+12 {{\rm e}^{-2 Br^2}}\pi  k^4r^2{\rho_0}^{2}+32  \left(\omega+\frac{1}{2} \right) \pi  Cr^2+k^4\sigma}}}}.
\end{eqnarray}
	
\begin{figure*}[thbp]
	\centering
	\includegraphics[width=5cm]{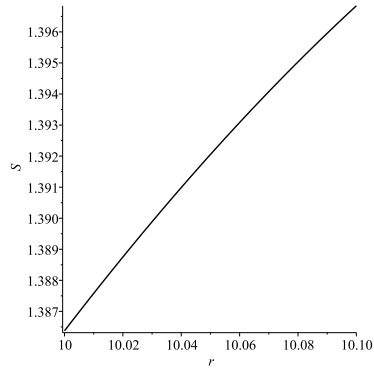}
	\caption{Variation of the entropy $(S)$ of the shell with respect to $r$.}
\end{figure*}

The shell entropy is  plotted along $r$ in Fig. 3 and exhibits a monotonic increase as we progress towards the exterior boundary of the shell. Also it can be noted that the entropy depends on the brane tension and bulk EOS parameter.

\subsection{Proper thickness}

The proper thickness of the finite thin shell can be calculated to have the form
\begin{eqnarray}\label{eq32}
\ell=& &\int_{R}^{R+\epsilon}\sqrt{e^\lambda}dr =\epsilon\sqrt{e^\lambda}\nonumber\\
& &= {\frac {\epsilon {k}^{2}\sqrt {\sigma  \left( 2 Br^2+1 \right)}}{\sqrt {12 {{\rm e}^{-2 Br^2}}\pi  k^4r^2{\rho_0}^{2}+8 \rho_0\pi  r^2 \left( k^4\sigma+2 A \left( 2 \omega+1 \right)  \right) {{\rm e}^{-Br^2}}+16 C\pi   \left( 2 \omega+1 \right) r^2+k^4\sigma}}}.
\end{eqnarray}

\begin{figure*}[thbp]
	\centering
	\includegraphics[width=5cm]{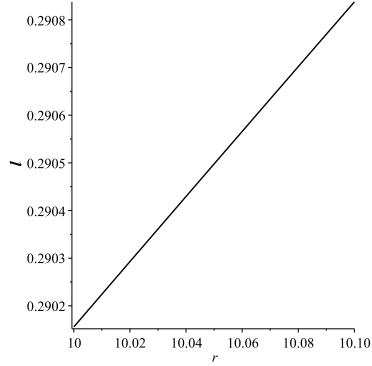}
	\caption{Variation of the entropy $(S)$ of the shell with respect to $r$.}
\end{figure*}

The proper thickness of the shell which has a strong dependence on the brane tension $\sigma$ and the EOS parameter $\omega$
concerning the bulk matter, is plotted as a function of the radial distance along the shell in Fig. 4. We observe that there is a linear dependence and the proper thickness increases.

\section{Boundary and matching conditions} \label{sec4}

When two space-times with different geometries and physical properties meet at a surface, then the junction conditions become important. The shell of the gravastar is different from the interior and exterior space-times in context of both the geometry and physical properties. So, following Lanczos~\cite{Lanczos}, we will use the well known Israel-Darmois junction conditions of GR~\cite{Darmois1927,Israel1966} to evaluate the components of the surface energy-momentum tensor expressed in terms of the extrinsic curvature. As the solutions of the EFE over the entire region are regular, there is analytic continuity of the metric function at the junction interface. The components, viz. the surface energy density and surface pressure are found to be of the form
\begin{eqnarray}\label{eq33}
   \Sigma & & =-\frac{1}{4\pi R}\bigg[\sqrt{e^\lambda}\bigg]_-^+ \nonumber\\& &
     =-\frac {1}{4\pi  R} \left(  \sqrt {1-{\frac {2 M}{R}}-\frac{\Lambda R^2}{3}}- \sqrt {1-\frac{4 R^2\pi   \rho_c}{3} \left( {\frac {2 \sigma+ \rho_c}{\sigma}} \right) -{\frac {16\pi   \left( A \rho_c+C \right) R^2}{k^4\sigma}}} \right),
	\end{eqnarray}

\begin{eqnarray}\label{eq34}
\mathcal{P} & =&\frac{1}{16\pi } \bigg[\bigg(\frac{2-\lambda^\prime R}{R}\bigg) \sqrt{e^{-\lambda}}\bigg]_-^+ \nonumber\\
&=&{\frac {-2 \Lambda {r}^{3}-3 M+3 r}{4\pi  r^2\sqrt {9-{\frac {18M}{r}}-3 \Lambda r^2}}}+\frac{\frac{4r\rho_c}{3}\frac{2 \sigma+\rho_c}{2 \sigma}+\frac{8 r (A \rho_c +C)}{k^4 \sigma}-\frac{16}{3 \pi r}}{2-\frac{4 r^2 \pi \rho_c(\sigma +\rho_c)}{2 \sigma}-\frac{16 \pi r^2 (A \rho_c +C)}{k^4 \sigma}}.
\end{eqnarray}

Here the symbol $\Lambda=\frac{6C}{\sigma}$ is the effective cosmological constant on the brane.

\begin{figure*}[thbp]
\centering
\includegraphics[width=5cm]{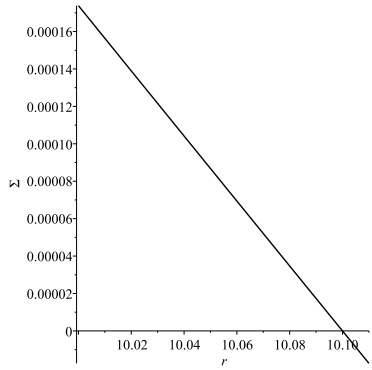}
\caption{Variation of the Surface energy density of the shell with respect to $r$.}
\end{figure*}

The surface energy density and surface pressure of the shell along the radial distance has been plotted in Figs. 5 and 6 respectively. The surface energy density $\Sigma$ is maximum at the interior boundary of the shell and decreases monotonically as we approach the shell-exterior interface and drops down to zero at this junction. On the contrary, the surface pressure $\mathcal P$ has a non-zero minimum at the interior boundary and increases monotonically until it reaches a maximum at the shell-exterior junction. Importantly, in this thin region of finite extent, both the surface energy and pressure parameters are non-negative, providing a good justification for the shell EOS, in accordance with the Mazur-Mottola prescription~\cite{Mazur2001,Mazur2004}. We also note from Fig. 5 that as soon as we exceed the radial limit for the exterior of the thin shell, the surface energy density of the shell flips sign to negative value, but it is of no physical significance as the exterior is the vacuum \textit{de Sitter} spacetime~\cite{Sengupta1}.

\begin{figure*}[thbp]
\centering
\includegraphics[width=5cm]{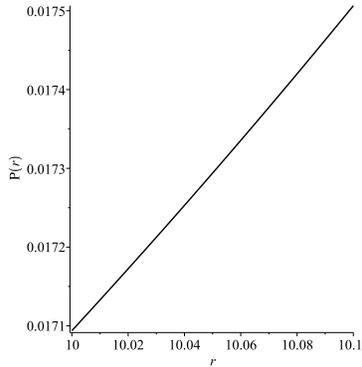}
\caption{Variation of the Surface pressure of the shell with respect to $r$.}
\end{figure*}

The mass of the thin shell (using Eq. (\ref{eq33})) is computed to have the form
\begin{eqnarray}
m_s&=&4\pi R^2 \Sigma \nonumber\\
 &=& R\left(  \sqrt {1-\frac{4 R^2\pi   \rho_c}{3} \left( {\frac {2 \sigma+ \rho_c}{\sigma}} \right) -{\frac {16\pi   \left( A \rho_c+C \right) R^2}{k^4\sigma}}} \sqrt {1-{\frac {2 M}{R}}-\frac{\Lambda R^2}{3}}\right).
\end{eqnarray}

Now the total mass of the gravastar can be calculated in terms of the mass of the thin shell as
\begin{equation}\label{20}
M_{Tot}=\frac{R}{18} \left( 9- \left( - {\frac {3 m_{{s}}}{R}}+ \sqrt{-24 R^2\pi  \rho_c \left(  {\frac {2 \sigma+\rho_c}{2\sigma}} \right) - {\frac {144\pi   \left( A\rho_c+C \right) R^2}{{k}^{4}\sigma}}+9} \right) ^{2}-3 \Lambda R^2 \right).
\end{equation}

The analytic continuation of the metric function $\lambda$ at the interior boundary of the shell provides us with the matching condition
\begin{eqnarray}\label{eq21}
	&&{\frac {12  \left( {{\rm e}^{- 2B R_1^2}} \right) ^{2}\pi  k^4{R_1}^{2}{\rho_0}^{2}+32  \left( \frac{1}{4} k^4\sigma+A \left(\omega+\frac{1}{2} \right)  \right) {R_1}^{2}\rho_0\pi  {{\rm e}^{-B R_1^2}}+32  \left( \omega+\frac{1}{2} \right) \pi  C{R_1}^{2}+k^4\sigma}{k^4\sigma  \left( 2 B{R_1}^{2}+1 \right) }}\nonumber\\
	&=&-\frac{8}{3} {R_1}^{2}\pi   \rho_c \left( \frac{1}{2} {\frac {2 \sigma+ \rho_c}{\sigma}} \right) -16 {\frac {\pi   \left( A \rho_c+C \right) {R_1}^{2}}{k^4\sigma}}+1.
\end{eqnarray}

It is also known that matching of the metric function $g_{tt}$ and $\frac{\delta g_{tt}}{\delta r}$ at the exterior boundary can help one to evaluate the constant model parameters. Therefore
\begin{equation}\label{eq22}
	1-{\frac {2m}{R_2}}-\frac{ \Lambda {R_2}^{2}}{3}={{\rm e}^{B{R_2}^{2}}}{K}^{2},
\end{equation}

\begin{equation}\label{eq23}
{\frac {2m}{{R_2}^{2}}}-\frac{2}{3} \Lambda R_2=2 B R_2{{\rm e}^{B{R_2}^{2}}}{K}^{2}.
\end{equation}

As we can see from Fig. 5, the surface energy density of the shell vanishes at the exterior boundary. We write this down as a condition to evaluate the external shell radius
\begin{equation}\label{eq24}
	-{\frac {1}{12 \pi  R_2} \left( \sqrt {9- {\frac {18m}{R_2}}-3 \Lambda {R_2}^{2}}- \sqrt {9-12 {R_2}^{2}\pi  \rho_c \left( {\frac {2 \sigma+ \rho_c}{\sigma}} \right) - {\frac {144\pi \left( A \rho_c+C \right) {R_2}^{2}}{k^4\sigma}}} \right) }=0.
\end{equation}

Now using the values of the known model parameters as $m = 3.750$,  $\omega = 0.63$, $R_1 = 10,\  R_2 = 10.1,\  k = 1,\  p_0 = 1,$  $\rho_c= 0.001$, $\Lambda = 0.00018$, $\sigma = 0.001$, we can obtain the constant parameters as $A = 0.001405658890, \ B = 0.085, \ C = -1.121931266*10^(-8),\ K = 0.006565223778$ by using the matching conditions given by Equations (32)-(34).

\section{Stability}

There are three main ways to check the stability of a compact object. We will perform these three methods of stability check for our gravastar model on the brane with the temporal metric potential being defined by the Kuchowicz function. First we obtain the surface redshift parameter for our model and check whether it is within the desired range for the gravastar to be stable. Secondly, we will check the validity of the energy conditions. Finally, we will check the validity of the Herrera's cracking condition for our gravastar model.

\subsection{Surface Redshift}

The surface redshift of the gravastar can be be obtained by using the formula
\begin{eqnarray}\label{eq35}
	Z_{s}&=&-1+\frac{1}{\sqrt {g_{\it tt}}} =-1+{\frac {1}{\sqrt {{K}^{2}{{\rm e}^{Br^2}}}}}.
\end{eqnarray}

\begin{figure*}[thbp]
	\centering
	\includegraphics[width=5cm]{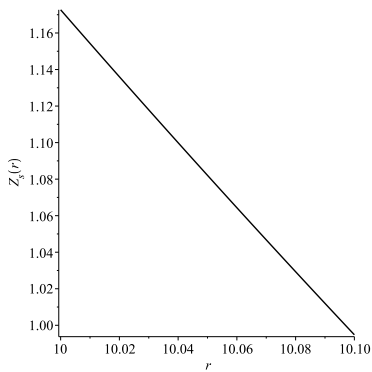}
	\caption{Variation of the Surface redshift of the shell with respect to $r$.}
\end{figure*}

The surface redshift along the radial distance across the shell has been plotted in Fig. 7. We know that in the absence of a cosmological
constant, the surface redshift $(Z_s)$ lies within the range $Z_s \leq 2$~\cite{Buchdahl1959,Straumann1984,Bohmer2006}. However, according to Bohmer and Harko~\cite{Bohmer2006}, the presence of a $\Lambda$- term relaxes the constraint on the surface redshift for a stable compact object to $Z_s\leq5$. As we see from the solution of the EFE in the space-time exterior to the gravastar that there is a dependence on the effective $\Lambda$-term present on the brane, so ideally for our gravastar to be stable, the surface redshift must be less than 5. However, from our plot we can note that $(Z_s)<2$ throughout the shell, which confirms the stability of our model.

\subsection{Energy conditions}

For a thin but finite spherical shell, composed of ultra-relativistic stiff fluid obeying the EOS $p=\rho$, the inequality conditions involving the components of the matter stress energy tensor must be valid within the context of modified brane gravity. As we know, on the brane the effective matter description holds consistent due to the corrections from the modified geometrical set-up. So, the energy conditions, viz. NEC (Null Energy Condition), WEC (Weak Energy Condition), SEC (Strong Energy Condition) and DEC (Dominant Energy Condition), all must be written down as: \\

NEC: $\rho^{eff} \geq 0,$      \\      \label{eq36}
WEC: $\rho^{eff}+p_r^{eff} \geq 0 , \rho^{eff}+p_t^{eff} \geq 0,$ \\ \label{eq37}
SEC: $\rho^{eff}+p_r^{eff}+2p_t^{eff} \geq 0,$         \\     \label{eq38}
DEC: $\rho^{eff}-|p_r^{eff}| \geq 0 , \rho^{eff}-|p_t^{eff}| \geq 0.$  \label{eq39}\\

\begin{figure}[!htp]
\centering
\includegraphics[width=5cm]{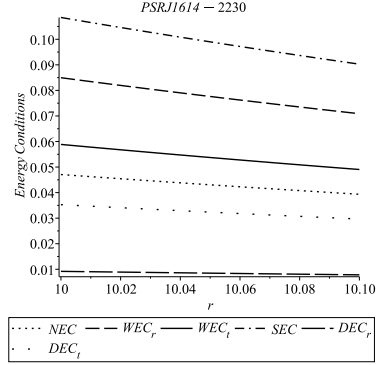}
\caption{Variation of the different energy conditions w.r.t. the radial coordinate $r$.}\label{pres.}
\end{figure}

As we can see from Fig. 8, where the variation of the essential combination of parameters required to form the energy conditions are shown along the radial distance $r$ across the shell, all the four energy conditions are valid, thus ensuring the stability of the compact object.

\subsection{Herrera's cracking condition}

If the radial forces are asymmetric in different regions of a spherically symmetric matter distribution, it leads to perturbation of the equilibrium matter distribution resulting in breaking or cracking. This asymmetry in the radial forces may appear from the fluid anisotropy. We will provide a final stability check for our gravastar model using the Herrera's cracking condition. As a consequence of causality, the sound speed squared inside the shell must obey $ 0\leq v^2_{ts} \leq 1$ and $ 0\leq v^2_{rs} \leq 1$ ~\cite{Herrera1992}. Also  as claimed by~\cite{Herrera1992} and~\cite{Andreasson2009}, it is essential for the radial sound $(v_{rs})$ speed to exceed the tangential sound speed$(v_{ts})$, implying that the inequality $|v^2_{rs}-v^2_{ts}| \leq 1$ must hold true for the matter distribution, so that no cracking occurs and the system is in stable configuration.

In our model, we get the parameters denoting the squared radial and tangential sound speeds respectively, as follows:
\begin{eqnarray}
&\qquad\hspace{-2.5cm}v^2_{rs}=\frac{dp^{eff}_r}{d\rho^{eff}}={\frac{3{{\rm e}^{-Br^2}}\rho_0 k^4+k^4\sigma+(4\omega+2) A}{{{\rm e}^{-Br^2}}\rho_0 k^4+k^4\sigma+6 A}},  \label{eq40} \\
&\qquad\hspace{-2cm}v^2_{ts}=\frac{dp^{eff}_t}{d\rho^{eff}}={\frac{3{{\rm e}^{-Br^2}}\rho_0 k^4+k^4\sigma-2 A ( \omega-1 ) }{{{\rm e}^{-Br^2}}\rho_0 k^4+k^4\sigma+6 A}}. \label{eq41}
\end{eqnarray}

\begin{figure}[!htp]
\centering
     \includegraphics[width=5cm]{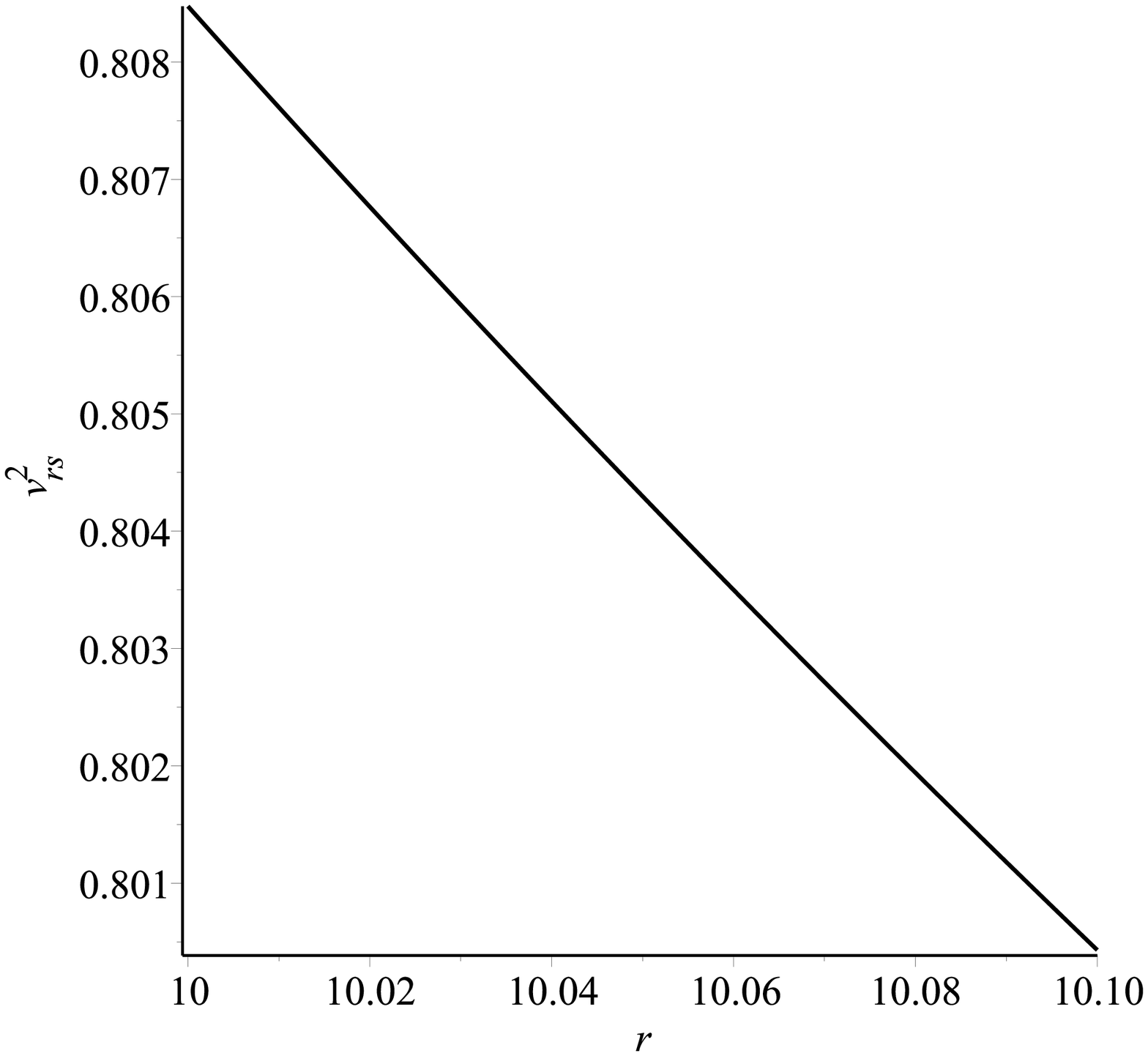}
     \includegraphics[width=5cm]{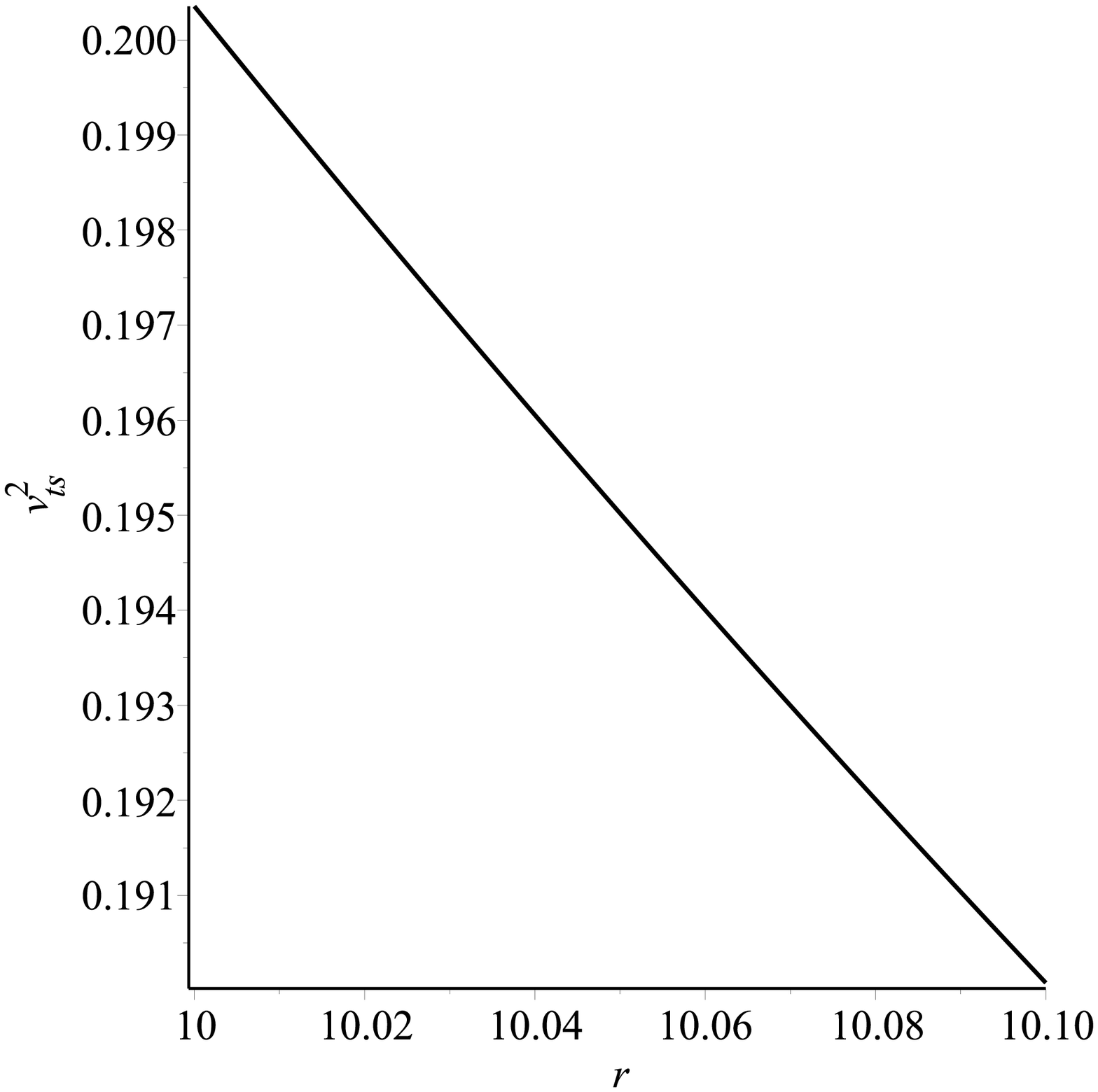}
\caption{Variation of $v^2_{rs}$ (left panel), $v^2_{ts}$ (right panel) w.r.t. the fractional
 radial coordinate $r/\Re$ for different strange star candidates.}\label{vel.}
\end{figure}

The causality conditions and the Herrera's cracking condition have been represented graphically in Figs. 9 and 10, respectively. As we can see in Figure 9, both $v^2_{rs}$ and $v^2_{ts}$ are less than $1$ throughout the shell. Both are maximum at the interior boundary and monotonically decrease until reaching the minimum at the surface of the gravastar. In Fig. 10, we see that  $|v^2_{rs}-v^2_{ts}|$ is always less than 1 at every point of the shell, thus satisfying the Herrera's cracking condition in addition to the causality conditions. The parameter increases linearly to a maximum at the surface, but the maximum value is also much less than unity. So, both the causality conditions and the cracking condition also hold true, thus confirming the stability of our gravastar model on the brane.

\begin{figure}[!htp]
\centering
     \includegraphics[width=5cm]{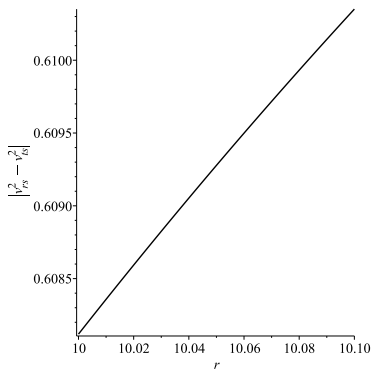}
\caption{Variation of $|v^2_{rs}-v^2_{ts}|$ w.r.t. the radial coordinate $r$.}\label{veld.}
\end{figure}

\section{Discussion and Conclusion}

In our earlier work~\cite{Sengupta1}, we have studied a gravastar model on the brane. In our present work we extend this study to test the effect of the Kuchowicz function as one of the metric potential describing the gravastar on the brane. Moreover, we have studied the construction of a wormhole on the brane with the redshift function being described by the Kuchowicz metric potential~\cite{Sengupta2}. As we can see, using Kuchowicz function has led to many interesting results in case of both the wormhole and gravastar on the brane.

The solution of the EFE in the interior is regular and depends on the brane tension. The matter density and pressure is constant as usual in the interior and the active gravitational mass is also dependent on the brane tension. Inside the shell the metric function obtained has an additional dependence on the bulk EOS parameter besides the brane tension. However, as in the case of our earlier model, the thin shell approximation is essential to solve the EFE in the shell analytically. We then move on to obtain the different physical parameters of the shell. As expected there is correction in all the parameters as a result of the higher dimensional effects coming into play. The entropy and proper thickness of the shell is found to depend on both the brane tension and the bulk EOS while the matter density, pressure and energy depends only on the brane tension. In our previous model the energy also had a dependence on the bulk EOS parameter. There is also another interesting difference from our previous model where the matter density and pressure was maximum at the surface of the gravastar but here we obtain that they have a non-zero minimum at the surface. This is much more acceptable as the matter density profile with decreasing values towards the surface seems to have greater stability. Also, we know that in the exterior just outside the surface the energy density and pressure must be zero. So, there was an abrupt drop in the energy density and pressure in our previous model but in the present case there is a smooth transition from slight non-zero minimum to zero. This provides a big advantage of using the Kuchowicz function.

From the Israel Darmois junction conditions\cite{Darmois1927,Israel1966}, with the help of the extrinsic curvature, we compute the surface energy density and pressure following the Lanczos\cite{Lanczos} prescription. We obtain a very interesting feature that the surface energy density vanishes at the surface of the gravastar which is again a desirable result for stability. However, the surface pressure increases monotonically before reache a maximum at the surface, while our previous gravastar model on the brane had a non zero minimum surface pressure. So, this is another effect of the Kuchowic function. A similar effect of the Kuchowicz potential can be observed when we study the surface redshift of our gravastar. In this model, in contrast to the monotonic increase of the surface redshift across the shell as obtained in our previous model, the surface redshift begins at a maximum value much less than 5 (as required for stability in presence of a cosmological constant) and even quite less than 2 (required for stability in absence of $\Lambda$), and has a minimum value of unity at the surface of the gravastar. The positive nature of the surface density and pressure along with the small value of surface redshift also shows the stability of our model.

To establish the stability of our present gravastar model further, we perform additional stability checks in the form of validity of the energy conditions of modified GR and also validity of the Herrera's cracking condition. As in RS brane gravity, we are provided with a geometrically modified description of matter on the 3-brane, so the energy conditions must be tested for the effective matter to accomodate the modifications due to the higher dimensional scenario. As we can see that the four energy conditions viz. the null, weak, strong and dominant, hold true within the shell of our gravastar. However, in case of our wormhole model with the Kuchowicz potential as the redshift function, we found the violation of the NEC for effective matter where it was essential in order to accomodate construction of traversable wormholes. Thus, both types of staic, spherically symmetric solutions obeying and violating the NEC can be accomodated on the brane with the temporal metric potential being described by the Kuchowicz function. This is an additional advantage of employing this metric function on the brane. We also find that our gravastar model passes another stability check such that radial sound speed is greater than the tangential sound speed, both having maximum value lesser than unity and minimum value at the surface of the gravastar, thus preventing any cracking effect in the shell matter composed of ultra-relativistic fluid arising from fluid anisotropy. This is an essential stability check as we have already found that pressure anisotropy is incorporated automatically in the brane gravastar from the non-local corrections on the brane arising from the projected Weyl tensor $E_{\mu \nu}.$

The important brane parameters that we have fixed to perform our entire analysis include the brane tension, the bulk EOS parameter and the effective cosmological constant on the brane. A small value of the brane tension $\sigma$ ensures that our analysis is valid in the high energy regime, where we actually look to operate involving the higher-dimensional corrections and providing a non-singular compact object model. In our work on wormhole\cite{Sengupta2}, we draw important constraint on the brane tension. In our present work we have used value of brane tension from within our constrained range to provide stable and successful model of gravastar. A positive value of $\omega$ less than one is also in good agreement with the Anti-deSitter($AdS_5$) description of the higher dimensional bulk space-time.

Thus, we make concluding remarks that we have substantially improved our previous model of the brane gravastar by using the Kuchowicz function as the temporal component of the metric potential, which we had previously done to construct a wormhole on the brane from ordinary matter. In a nutshell, the most important distinguishing features from our previous model are- the vanishing of the surface energy density and the minimum of matter density and pressure values at the surface of the gravastar, the surface redshift being within the stability criteria and also vanishing at the surface of the gravastar, the non-dependence of the shell energy on the bulk EOS parameter, successful stability check in the form of validity of NEC and other energy conditions to see whether both classes- NEC obeying and NEC violating  static, spherically symmetric solutions exist on the brane and final stability check in form of Hererra's cracking condition which is extremely relevant for the ultra-relativistic shell to avoid possible instabilities due to cracking. The values of the essential physical model parameters viz. brane tension $\sigma$ and bulk EOS parameter $\omega$ that we have used for our analysis is also well within the obtained range of constrained values. Therefore, the Kuchowicz metric potential is very efficient for describing compact objects on the brane at the high energy regime, including regular gravastar solutions and regular traversable wormholes.

However, finally we would like to briefly provide a quick review of the observational status in this field which is as follows:

Indirect methods for the possible detection of gravastars have been discussed in~\cite{Sakai2014,Kubo2016,Cardoso1,Cardoso2,Abbott2016,Chirenti2016}. Sakai et al.~\cite{Sakai2014} suggested the analysis of gravastar shadows to be considered a tool in gravastar detection. Also the use of gravitational lensing phenomenon can be made~\cite{Kubo2016} where the authors have claimed to find gravastar microlensing effects distinguishable in maximal luminosity from that of identical mass black holes. Cardoso et al.~\cite{Cardoso1,Cardoso2} have interpreted the ringdown signal of $GW~150914$~\cite{Abbott2016}, detected by interferometric LIGO detectors, to be generated by horizonless compact objects, and as we know, gravastars are one of the leading candidate for such objects. However, there is no final confirmation on this matter yet~\cite{Chirenti2016}. The brane-bulk geometry can be applied to the interpretation of $GW~170817$ event from the LIGO/Virgo detectors and its associated EM couterpart~\cite{Visinelli}.Also recently, the observation of the dark shadow of  M87$^{\ast}$ has been successfully explained within the context of the Randall-Sundrum brane gravity~\cite{Vagnozzi}.

\section*{Acknowledgement}
SR gratefully acknowledges support from the Inter-University Centre for Astronomy and Astrophysics (IUCAA), Pune, India under its Visiting Research Associateship Programme. The present research was granted by the DST-WB, Higher Education Department of Science, Technology and Biotechnology under the Project ``Studies on compact stellar objects", Memo No. 171(Sanc.)/ST/P/S\&T/16G-17/2018.

\end{document}